\documentstyle[prl,twocolumn,epsf,eqsecnum,aps]{revtex}

\begin{document}
\draft
\title{Acoustic phonon exchange, attractive interactions, and the
Wentzel-Bardeen singularity in single-wall nanotubes}
\author{A. De Martino and R. Egger}
\address{Institut f\"ur Theoretische Physik, 
Heinrich-Heine Universit\"at,
D-40225 D\"usseldorf, Germany}
\date{Date: \today}
\maketitle
\begin{abstract}
We derive the effective low-energy theory for interacting electrons 
in metallic single-wall carbon nanotubes taking into account acoustic 
phonon exchange within a continuum elastic description.
In many cases, the nanotube can be described as a standard Luttinger
liquid with possibly attractive interactions.  
We predict surprisingly strong attractive interactions for
thin nanotubes.  Once the tube radius reaches a critical value
$R_0 \approx 3.6\pm 1.4$~\AA, the 
Wentzel-Bardeen singularity is approached, accompanied by
strong superconducting fluctuations.  The surprisingly large $R_0$ indicates that
this singularity could be reached experimentally.
We also discuss the conditions for a Peierls transition due to acoustic phonons.
\end{abstract}

\narrowtext

\section{Introduction}

Superconductivity in carbon nanotubes has lately generated a lot of excitement.
Experimental observations include an unexpectedly strong proximity effect
in single-wall nanotubes (SWNTs) \cite{morpurgo,kasumov},
 fluctuation superconductivity in ultrathin SWNTs \cite{tang}, 
and intrinsic BCS-type behavior in ``ropes'' containing many SWNTs \cite{kociak}. 
These experiments have in turn led to a number of theoretical papers 
\cite{gonzalez,caron,leon,byczuk}.  In particular, the authors of
Refs.~\cite{gonzalez,caron} argue that phonon exchange  holds responsible for
some attractive electron-electron interaction.  
The value or even order of magnitude of the electron-phonon coupling was however
left open or simply extracted by fitting experimental curves.  
In this work, we wish to fill this gap
and provide a quantitative analysis of phonon-mediated
retarded electron-electron interactions in a metallic SWNT. (The theory
also holds qualitatively for a heavily doped semiconducting SWNT.)
Below we include the usual (repulsive) Coulomb interactions 
in the effective low-energy theory
valid at energies $|E|<v_F/R$, with Fermi velocity $v_F$ and SWNT radius $R$.
We assume that the SWNT is doped away from the charge neutrality point, 
$E_F\neq 0$, i.e., an incommensurate situation,
which is normally encountered in practice.
Then electron-phonon and electron-electron Umklapp processes can be neglected.

Besides mediating possibly 
attractive interactions, the effect of phonons on the electronic
system could in principle lead to two instabilities, namely the Peierls 
transition \cite{gruner} and the Wentzel-Bardeen (WB) singularity \cite{wb}.
All these issues can be investigated by integrating out the phonons and 
studying the electron system alone, since the relevant backscattering
electron-phonon interactions are weak \cite{voit}.
This is the approach taken in our paper.
Our treatment of phonons expands on recent work by Suzuura and Ando
\cite{suzuando}, where the field theory for {\sl acoustic phonons} 
and their coupling to electrons in SWNTs was derived.
Optical phonons are ignored in this theory as well as in ours. This should
create no major problem for the issues at stake here, since optical phonons
do not produce sizeable attractive interactions nor a WB singularity.
Their relevance for the Peierls transition is discussed briefly below.
The theory of Ref.~\cite{suzuando}
is analytically tractable and, moreover,
appears to be in excellent agreement with
local-density-functional calculations even for ultrathin  ($R\approx 2$~\AA)
nanotubes \cite{liu}. 
For instance, we find that the frequency of the breathing mode computed in 
Ref.\cite{liu}
coincides to within $4\%$ with the value extracted from the continuum theory of 
Ref.\cite{suzuando}.  Furthermore, the predicted value for the twist velocity $v_T$
below is consistent with Refs. \cite{kane,peierls}.
This is fortunate since previous estimates for the electron-phonon coupling
\cite{benedict,huang} were only qualitative in nature 
and did not even agree on orders of magnitude.  
For related theoretical work on acoustic phonons and electron-phonon interactions
in SWNTs, see Ref.\cite{mahan}.

As we show in detail below, the phonon-mediated effective
attraction among electrons is very strong for sufficiently thin nanotubes.
The dominant phonon exchange processes involve two phonon modes, namely
the stretching and the breathing mode.  The stretching mode has a linear
dispersion characterized by a velocity $v_S$, while the breathing mode
has a finite frequency $\omega_B$ at long wavelengths.   
Following general arguments \cite{voit,car}, for a large part of the relevant
parameter space, the influence of the stretching mode is negligible.  Then
the system can be described as a {\sl Luttinger liquid}\ (LL) with
standard interaction parameter $K_{c+}$ in the charged channel \cite{egger}, 
where $K_{c+}=1$ for noninteracting electrons.  Phonon exchange involving the
breathing mode implies 
\begin{equation}\label{geff}
K_{c+} = \frac{K^{0}_{c+}}{\sqrt{1- (K^{0}_{c+})^2R_B/R}},
\end{equation}
with $R_B= 2.4 \pm 0.9$~\AA.
Here $K^{0}_{c+}\leq 1$ 
characterizes the repulsive Coulomb interactions,
where $K^{0}_{c+}\approx 0.2$ for unscreened interactions \cite{egger}.
Notice that without external screening, 
phonon exchange is not relevant, and $K_{c+}\approx K^{0}_{c+}$. 
However, once  Coulomb interactions are externally screened off, 
e.g. by nearby gate electrodes, we get $K^{0}_{c+}\approx 1$ and
Eq.~(\ref{geff}) predicts surprisingly strong {\sl attractive interactions}
($K_{c+}>1$). For a $(10,10)$ armchair tube with $R=6.79$~\AA, the LL parameter 
is then $K_{c+}\approx 1.3$, i.e. significantly larger than the noninteracting value.
These attractive interactions get stronger for thinner nanotubes.

As we discuss below, the neglect of the retarded interaction due to
the stretching mode, leading to a (four-channel) 
LL with interaction parameter $K_{c+}$ given by Eq.~(\ref{geff}) and interaction
parameters $K_\lambda=1$ in the three neutral channels (see below),
is valid provided $R$ is sufficiently large compared to the 
``critical  radius'' $R_c= (K^0_{c+})^2 R_0$, where $R_0$ is slightly larger than $R_B$, 
$R_0 = 3.6 \pm 1.4$\AA. Only in the immediate vicinity of this radius, 
the retarded interaction mediated by the stretching
phonon mode has to be kept explicitly.
This critical radius defines the location of the so-called Wentzel-Bardeen (WB)
singularity \cite{wb}.  For $R\leq R_c$, our model breaks down, and a transition to a 
phase-separated state is expected.  
Previous work by Loss and Martin \cite{loss}
studied the WB singularity in detail for one-dimensional (1D) systems, but 
a concrete physical realization seemed out of reach due to the 
small ratio between sound and Fermi velocity in all systems considered so far. 
We argue below that SWNTs with screened-off interactions may
offer the possibility to reach this elusive singularity,
which has never been observed experimentally.
This is remarkable since the ratio between sound and Fermi
velocity is still small.  However, in a SWNT the stretching and the breathing
phonon modes collaborate in driving the system towards the WB 
singularity in a very efficient manner.

Moreover, the closeness to the WB singularity holds responsible
for rather strong superconducting fluctuations.  From our theory, 
estimates for the attractive interactions generated via phonon 
exchange are obtained. To describe most of the issues raised by the
experiments of Refs.\cite{tang,kociak} in a quantitative way, 
it will however be important to also analyze the effect of various 
inter-tube couplings in SWNT ropes or crystals in more detail.  
Employing a mean-field approximation for Josephson
couplings, the standard 1D Ginzburg-Landau description can be obtained \cite{schulz}.
In this paper, we focus on the phonon-mediated electron-electron interaction within a 
given nanotube, and postpone a full discussion of experimentally relevant quantities
to a future publication \cite{future}.  

The present paper also sheds light on the recent discussion about the
possibility of a Peierls transition in SWNTs \cite{peierls,p1,peierls2,peierls3}.
Notably, even in the incommensurate situation studied here, in principle
a Peierls transition can occur \cite{senna}, although the resulting 
pseudogap is expected to be wiped out by quantum fluctuations \cite{voit}.
Within a mean-field approximation,
the Peierls transition due to optical phonons has been studied in 
Refs.\cite{p1,peierls2,peierls3}, resulting in a Kekul{\'e}-type modulation.
Since we only treat acoustic phonons, 
we have nothing to say about this issue.
However, recent theoretical work  including acoustic phonons has
pointed out the possibility of a sizeable Peierls gap in thin armchair SWNTs 
\cite{peierls}.  This Peierls modulation was found to correspond to a twist
distortion of the SWNT.  Below we study arbitrary metallic SWNTs, and indeed
confirm that the relevant Peierls distortion for an armchair SWNT would
correspond to a tube twisting.  In addition, we find the proper mode for any
metallic SWNT.  
Our findings follow from analyzing backscattering phonon exchange, which operates
at the doping-dependent wavevector $2q_F$, where $q_F=|E_F|/v_F$.  
Based on our essentially exact results, we do not expect that the  
Peierls transition predicted in Ref.\cite{peierls} is observable,
at least not for an individual SWNT away from the neutrality point.
According to our theory, a Peierls gap in the 1D sense \cite{voit}
could only be possible for very thin SWNTs and under rather stringent 
conditions.

The outline of the paper is as follows.
In Sec.~\ref{sec2} we briefly summarize the elastic continuum theory for
phonons and the electron-phonon interaction. The resulting effective 
electron-electron interaction generated by phonon exchange is derived in 
Sec.~\ref{sec3}.  The dominant forward scattering process, the Luttinger liquid
with attractive interactions, and the 
Wentzel-Bardeen singularity are discussed in
Sec.~\ref{sec4}.  The remaining phonon exchange terms and the
Peierls transition is studied in Sec.~\ref{sec5}.
Some details concerning Sec.~\ref{sec5} have been transferred to an appendix.
Finally, we discuss the relevance of inter-tube phonon
exchange and conclude in Sec.~\ref{sec7}.  

\section{Acoustic phonons and electron-phonon coupling}
\label{sec2}

Here we shall always assume a suspended (free-standing) tube.  With regard to the
important phonon modes coupling to the low-energy electrons, this should be
an accurate assumption both for SWNTs embedded in a zeolite matrix \cite{tang}
or in a rope \cite{kociak}.  However, modifications may arise for individual SWNTs
on a substrate, where phonon modes could be pinned. 
 Let us then briefly summarize the main results of Ref.~\cite{suzuando},
where acoustic phonons 
in SWNTs are described by an elastic continuum theory.
The Euclidean action is (we put $\hbar=1$)
\begin{equation} \label{phac}
S = \int  d\tau   d^2r  \; \left \{ (M/2)\dot {\bf u}^2 +
 U[{\bf u}] + U_c[{\bf u}] \right\},
\end{equation}
where ${\bf u}^T(\vec r)=(u_x,u_y,u_z)$ is the displacement field and $\vec r=(x,y)$.
The displacement field in this approximation does not discriminate among 
the graphite sublattices, and hence will not describe optical phonons.
The coordinate system
is chosen as follows. The $y$-axis points along the tube, $0\le x\le 2\pi R$ around
the circumference, and $z$ denotes the direction perpendicular to the tube surface.
The elastic potential energy density in Eq.~(\ref{phac}) is given by
\[
U[{\bf u}] = \frac12 \left( B(u_{xx}+u_{yy})^2 + 
\mu [ ( u_{xx} - u_{yy} )^2 + 4 u_{xy}^2 ]  \right),
\]
where the strain tensor is defined by the relations $u_{yy}   = \partial_y u_y$, 
$2 u_{xy}  = \partial_y u_x+ \partial_x u_y$, 
and, due to the cylindrical geometry, 
$u_{xx} = \partial_x u_x + u_z/R$.
Furthermore, the curvature-induced potential energy density is
\[
U_c[{\bf u}] = (a^2 \Xi/2) 
[ ( \partial_x^2 + \partial_y^2 + R^{-2}) u_z ]^2.
\]
In the above expressions, $a=2.46$~\AA~is the lattice constant, and 
the carbon mass per unit area is
$M=3.80\times 10^{-7}$~kg/m$^2$. The bulk modulus $B$ and the
shear modulus $\mu$ are accurately known for bulk graphite,
\begin{eqnarray} \label{matpar}
B/M & = & 2.90 \times 10^8 \, {\rm m}^2/{\rm sec}^2 ,\\  \nonumber
\mu/M  &=& v_T^2= 1.51 \times 10^8 \, {\rm m}^2/{\rm sec}^2. 
\end{eqnarray}
Finally, the  normal-force constant is $\Xi/M  = 6.19 \times 10^6 \, {\rm m}^2/{\rm sec}^2$.
Since $\Xi \ll B, \mu$ and the curvature-induced terms are additionally suppressed in 
the long-wavelength limit, 
terms related to $\Xi$ are neglected in what follows.
It is then convenient to switch to Fourier space,
\[
{\bf u}(\tau,\vec r) = \frac{1}{\beta L} \sum_{\omega_n,q_x} 
\int \! \frac{dq_y}{2\pi} e^{-i( \omega_n \tau - \vec q \cdot \vec r)} 
{\bf u}(\omega_n,\vec q),
\]
where $\vec q =(q_x,q_y)$, $\omega_n = 2\pi n/\beta$ with $\beta=1/k_B T$ 
are Matsubara frequencies, $L=2\pi R$ is the circumference of the tube, and 
$q_x=n/R$ with integer $n$ denotes the discrete transverse momenta. 
The action (\ref{phac}) then reads  
\begin{equation}
S = \frac{1}{2 \beta L} \sum_{\omega_n,q_x} \int 
\frac{dq_y}{2\pi} \; {\bf u}^{T} (-\omega_n,-\vec q) 
\underline{A} (\omega_n,\vec q) {\bf u}(\omega_n,\vec q),
\end{equation}
where $\underline{A}$ is given by the matrix
\[
\left(\begin{array}{ccc}
M\omega_n^2+\Delta q^2_x + \mu q_y^2 & B q_x q_y & -i\Delta \frac{q_x}{R}\\
B q_x q_y & M\omega_n^2+\Delta q^2_y + \mu q_x^2 & -i(B-\mu) \frac{q_y}{R}\\
i \Delta \frac{q_x}{R} & i(B-\mu) \frac{q_y}{R}  &
M\omega_n^2+\frac{\Delta}{R^2} 
\end{array}\right)
\]
with $\Delta=B+\mu$.

Our strategy is to integrate out phonons, which is valid unless the
electron-phonon coupling is very strong and/or soft phonon modes are present, 
see Ref.\cite{voit}.  These requirements are met in our case.
In order to proceed, we then need the inverse of $\underline{A}$, whose
general form is straightforward to obtain but not required here. 
The only  transverse momentum relevant for the effective 
interaction is $q_x=0$, since only the corresponding 
electronic states have to be kept at low energy scales.
Therefore we focus on this special limit henceforth.  
Putting $q_y=q$ and $q_x=0$,  the matrix $\underline{A}$ reads
\[
\underline{A}=\left(\begin{array}{ccc}
M\omega_n^2 + \mu q^2 & 0 & 0\\
0 & M\omega_n^2+  \Delta q^2 & -i(B-\mu) q/R \\
0 & i(B-\mu) q/R  &
M\omega_n^2+\Delta/R^2
\end{array}\right) ,
\]
with eigenvalues $(i=1,2,3)$
\begin{equation}\label{lambdas}
\lambda_{i}  = M \left[ \omega_n^2 + \epsilon^2_{i}(q) \right].
\end{equation}
The inverse matrix is 
\begin{equation} 
\label{invA}
\underline{A}^{-1} = \left(
\begin{array}{ccc}
\frac{1}{\lambda_1} & 0 & 0 \\
0 & \frac{M\omega_n^2 R^2+\Delta}{\lambda_2 \lambda_3 R^2}  
 & \frac{i(B-\mu) q}{\lambda_2 \lambda_3 R}  \\
0 & - \frac{i(B-\mu)q}{\lambda_2 \lambda_3 R} &  
\frac{M\omega_n^2+\Delta q^2}{\lambda_2 \lambda_3}  
\end{array}
\right) .
\end{equation}
 Here $\epsilon^2_1 (q) = v_T^2 q^2$ with $v_T$ in Eq.~(\ref{matpar}),
and 
\begin{equation}
 \epsilon^2_{2,3}(q) = (\omega_B^2/2)[1+q^2 R^2\mp E(q)]
\end{equation}
 with 
\begin{equation}\label{defb}
E(q)= [1+ 2 q^2 R^2 (1  -8B\mu/\Delta^2)+q^4 R^4 ]^{1/2}.
\end{equation}
In the long-wavelength limit, $qR \ll 1$, further simplifications 
are possible.  Although the theory is valid also for larger $qR$, we shall
make this inessential approximation whenever appropriate.  
To  lowest order in $qR$, we obtain for the three eigenmodes
\begin{equation}\label{eps}
\epsilon_1^2(q) = v_T^2 q^2, \quad \epsilon_2^2(q)=v_S^2 q^2, \quad \epsilon_3^2(q)=\omega_B^2,
\end{equation}
where 
\begin{equation}
v^2_S =  4 B\mu / \Delta M, \quad \omega^2_B= \Delta/MR^2.
\end{equation}
The eigenmodes corresponding to $\lambda_{1,2,3}$ 
are the twisting, stretching and
breathing phonon modes, respectively.  The sound velocities for the twisting and
stretching modes are $v_T$ and $v_S$, and the breathing mode frequency is $\omega_B$.
{}From Eq.~(\ref{matpar}) we obtain the following estimates
\begin{eqnarray}
\nonumber
v_T &=& 1.23 \times 10^4~{\rm m/sec} ,\\
\label{matpar2}
v_S &=& 1.99 \times 10^4~{\rm m/sec} ,\\
\nonumber
\hbar\omega_B &=& \frac{0.14}{R}~{\rm eV \AA} .
\end{eqnarray}
Note that both $v_S$ and $v_T$ are much smaller than the Fermi 
velocity, $v_F=8\times 10^5$~m$/$sec. 
The stretching mode corresponds to the longitudinal acoustic (LA) phonon,
while the twisting and breathing ones correspond to transverse acoustic 
(TA) phonons.

Next we address the coupling of the phonon  modes to electrons.
The relevant electron states for a metallic SWNT are spinors $\psi_{p\alpha\sigma}(y)$ 
\cite{egger}, where $p=\pm$ denotes the sublattice (the honeycomb lattice has two
basis atoms), $\alpha=\pm$ labels the two distinct K points, 
and $\sigma=\pm$ the spin.
Suppressing the $\alpha,\sigma$ indices, the electron-phonon coupling corresponds
to Hamiltonian densities ${\cal H}^{K,K'}_{e-ph}$ for the two K points, 
\begin{equation}
{\cal H}^{K}_{e-ph} = \left(
\begin{array}{cc}
V_1 & V_2 \\
V_2^* & V_1 
\end{array} 
\right) , \quad 
{\cal H}^{K'}_{e-ph} = \left(
\begin{array}{cc}
V_1 & V^*_2 \\
V_2 & V_1 
\end{array} 
\right) .
\end{equation}
Here
the {\sl deformation potential} is
\begin{equation}  \label{defo}
V_1 = g_1 (u_{xx} + u_{yy}) ,
\end{equation}
where $g_1\approx 16$~eV in bulk graphite \cite{sugi}.
For a 2D graphite sheet this value should 
be multiplied by $3/2$ \cite{suzuando},
and we therefore take the estimate $g_1\approx 20-30$~eV.
The off-diagonal terms in sublattice space arise from a
bond-length change,
\begin{equation} \label{v2}
V_2 = g_2 e^{3 i \eta } (u_{xx}-u_{yy} + 2i u_{xy} ),
\end{equation}
where $\eta$ is the chiral angle, with $\eta=0$ for zig-zag tubes and
$\eta=\pi/6$ for armchair tubes.  
The respective coupling constant is $g_2 \approx 1.5$~eV \cite{suzuando}. 
Although both the free electron and the free phonon low-energy theory do not
depend on chirality, the $V_2$ term in the electron-phonon coupling does.

\section{Interactions from phonon exchange}\label{sec3}

We can now integrate out the phonons and obtain an effective action 
for the electrons that includes the phonon-mediated retarded interaction.
To that purpose, we evaluate the Gaussian path integral for the phonons, 
\begin{eqnarray}\label{ph}
&& \int {\cal D}{\bf u} \exp
\Bigl\{ -\frac{1}{2L} \int [dq]  \Bigl[ {\bf u}^T(-\omega_n,-q) \underline{A}(\omega_n,q)
{\bf u}(\omega_n,q) \nonumber \\  
&& + 2 {\bf V}^T(-\omega_n,-q) {\bf u}(\omega_n,q) \Bigr] \Bigr\} = e^{-S_{e-ph}},
\end{eqnarray}
with the notation
\begin{equation}\label{meas}
\int [dq] =  \frac{1}{\beta} \sum_{\omega_n} \int_{-\infty}^\infty \frac{dq}{2\pi}.
\end{equation}
The result $S_{e-ph}$ defines the phonon-mediated contribution to the electronic action.
In Eq.~(\ref{ph}), we use the vector
\begin{eqnarray}
\label{vi} 
{\bf V}(\omega_n,q) &=& g_1 \left[ \sum_\alpha \rho_{\alpha}(\omega_n,q) \right]
{\bf M}(q) \\
\nonumber &+& \frac{g_2}{2} e^{3i\eta}[J_{1+}(\omega_n,q)+J_{2-}(\omega_n,q)]  
{\bf N}_+(q) \\
\nonumber & +&  
\frac{g_2}{2} e^{-3i\eta} [J_{1-}(\omega_n,q) + J_{2+}(\omega_n,q)] {\bf N}_-(q).
\end{eqnarray}
Fourier transforming $V_1$ and $V_2$ in Eqs.~(\ref{defo}) and (\ref{v2}) 
specifies the auxiliary vectors
${\bf M}^T(q)  =  (0, -iq, 1/R)$,  
${\bf N}^T_+(q) =  ( q, iq, 1/R)$, and
${\bf N}_-(q) =  {\bf N}^\ast_+ (-q)$. 
Using fermionic Matsubara frequencies 
$p_n=(2n+1)\pi/\hbar \beta$ 
and analogous integration measure $\int [dp]$, electronic densities are
\[
\rho_\alpha (\omega_n,q) = \int [dp] \ \overline \psi_{\alpha}(p_n,p) 
\psi_{\alpha}(p_n+\omega_n,p+q).
\]
Similarly, electronic currents  are defined as
\begin{equation} \label{jap}
J_{\alpha\pm} (\omega_n,q) = \int [dp] \ \overline \psi_{\alpha}(p_n,p)  \tau_\pm 
\psi_{\alpha}(p_n+\omega_n,p+q) .
\end{equation}
The fermionic (Grassmann) fields represent spinors, where we have suppressed 
the sublattice
and spin indices, and Pauli matrices
$\tau_\pm= \tau_1 \pm i \tau_2$ act in sublattice space.
The integration over phonons is now straightforward, leading to the
phonon-mediated action contribution
\begin{equation} \label{eleffac}
S_{e-ph}= - \frac{1}{2L}
\int [dq] \  {\bf V}^T(-\omega_n,-q) {\underline A}^{-1}(\omega_n,q) 
 {\bf V}(\omega_n,q) .
\end{equation}
This form will be analyzed in the remainder of the paper. 

To make further progress, it is convenient to invoke the 
Abelian bosonization scheme adapted to SWNTs  \cite{egger}.
To keep the paper self-contained, we briefly review the
ingredients needed for our subsequent discussion.
This requires first a unitary transformation in sublattice space, 
which transforms the free-electron 
Hamiltonian into the standard form containing right- and left-moving  
fields,
$\psi_{p\alpha\sigma} = \sum_{r=R/L=\pm} U_{pr} \tilde{\psi}_{r\alpha\sigma}$,
where 
\[
U=\frac{-e^{-i\pi/4}}{\sqrt{2}} 
\left( \begin{array}{cc}
1 & 1 \\
i & -i
\end{array} \right).
\]
Under the unitary transformation, we have to replace
$\tau_2 \to \tau_3$, $\tau_3\to \tau_1,$ and $\tau_1\to \tau_2$.
We then work in the transformed picture, and omit the tilde in the field
operators henceforth.
The bosonization formula for the electron operator is \cite{egger}
\begin{equation}\label{boson}
\psi_{r\alpha\sigma} (y) = \frac{\eta_{r\alpha\sigma}}{\sqrt{2\pi a}} 
\exp\{i( r q_F y + \alpha k_F y + r \sqrt{4\pi} \phi_{r\alpha\sigma}) \} ,
\end{equation}
where $\eta_{r\alpha\sigma}$ are Klein factors, 
$k_F$ is the $y$-projection of the K point, and $q_F=|E_F|/v_F$ depends on the doping 
level $E_F$.
The chiral fields $\phi_{r\alpha\sigma}$ are related to a pair of dual nonchiral
boson fields,
$\phi_{\alpha\sigma} =\phi_{R\alpha\sigma} + \phi_{L\alpha\sigma}$ and
$\theta_{\alpha\sigma}=\phi_{R\alpha\sigma} - \phi_{L\alpha\sigma}$.
Switching to  total/relative charge/spin fields,
$\phi_{c\pm} = \frac{1}{2} \sum_{\sigma} 
(\phi_{1\sigma}\pm\phi_{2\sigma})$ and $\phi_{s\pm} = \frac{1}{2} \sum_{\sigma} 
\sigma (\phi_{1\sigma}\pm\phi_{2\sigma})$,
and similarly for the dual fields, the electronic action 
including repulsive forward-scattering Coulomb interactions but without phonon exchange
is described by a four-channel ($\lambda=c+,c-,s+,s-)$
Luttinger liquid \cite{egger},
\begin{equation} \label{bosac}
S_0 = \sum_\lambda \frac{1}{2} \int [dq] 
\frac{|\phi_\lambda(\omega_n,q)|^2}{ u^0_{\lambda} K^0_{\lambda}}
 (\omega_n^2+(u^0_\lambda)^2 q^2) .
\end{equation}
We  ignore the tiny electron-electron 
backscattering processes \cite{egger} in this paper. 
The velocities are $u^0_\lambda=v_F/K^0_\lambda$, where $K^0_\lambda=1$
in all neutral channels ($\lambda\neq c+$).
For the charged channel, $u^0_{c+}=v_F/K^0_{c+}$, 
where the Luttinger liquid parameter $K^0_{c+}$ depends
on the electron-electron interaction strength; $K^0_{c+}=1$ for the noninteracting
problem.  In the presence of unscreened repulsive interactions, $K^0_{c+}\approx 0.2$.
In the rest of the paper we shall use the symbols $u_\lambda, K_\lambda$ 
without superscript to denote the values of the effective parameters
renormalized by the phonon exchange.  
After this short review of bosonization for SWNTs, 
we now analyze the various terms in the phonon-mediated action $S_{e-ph}$ in 
Eq.~(\ref{eleffac}) using bosonization.

\section{Attractive interactions and Wentzel-Bardeen singularity}\label{sec4}

To begin with, there is a dominant forward-scattering term 
coupling to the total electronic density, which in bosonized form reads
$\rho(y)=(2/\sqrt{\pi}) \partial_y \phi_{c+}$.
The respective contribution to Eq.~(\ref{eleffac}) is
\begin{equation} \label{fs}
S_f = -\frac{g_1^2}{2L} \int [dq] \ \rho(-\omega_n,-q) D_f(\omega_n,q) 
 \rho(\omega_n,q),
\end{equation}
with the inverse propagator
\begin{eqnarray*}
D_f(\omega_n,q) &=&{\bf M}^T(-\omega_n,-q) {\underline A}^{-1}(\omega_n,q) 
{\bf M}(\omega_n,q)\\
&=& \frac{ M\omega_n^2(1+q^2R^2) + 4\mu q^2}{\lambda_2\lambda_3 R^2} .
\end{eqnarray*}
This is the only phonon exchange contribution of order $g_1^2$. Due to
the large deformation potential [as compared to the coupling $g_2$], it  is 
indeed expected to be the dominant one.
 
Notice that the twisting mode, corresponding to the eigenvalue 
$\lambda_1$, does  not contribute to $D_f$. In order to make 
explicit the contributions from the other two phonon branches,
we can rewrite $D_f$ as follows,
\begin{equation}
D_f (\omega_n,q) = \frac{1}{M R^2}
\left[ \frac{Z_2(q,\mu)}{\omega_n^2 + \epsilon_2^2(q)} + 
\frac{Z_3(q,\mu)}{\omega_n^2 + \epsilon_3^2(q)}
\right],
\end{equation}
where we use the functions
\begin{equation}\label{coefz2}
Z_{2,3}(q,\Lambda) = \pm\frac{R^2}{\Delta E(q)}\left[ 
-(1+q^2R^2) M \epsilon^2_{2,3}(q) + 4 \Lambda q^2
\right] ;
\end{equation}
$E(q)$ is defined in Eq.~(\ref{defb}). 
The contribution (\ref{fs}) thus only operates
in the charged channel, there is no effect on neutral channels. Including 
the action $S_0$ in Eq.~(\ref{bosac}),
the electronic action in the charged channel then takes the form
\begin{eqnarray} \label{finfa}
S[\phi_{c+}] &=& \frac{1}{2 v_F} \int [dq] 
| \phi_{c+}(\omega_n,q)|^2 \Bigl [ \omega_n^2 
+ (u^0_{c+})^2 q^2 \\ \nonumber
& - &  \gamma q^2 \left( \frac{Z_2(q,\mu)}{\omega_n^2 + 
\epsilon_2^2(q)} + \frac{Z_3(q,\mu)}{\omega_n^2 + 
\epsilon_3^2(q)}
\right) \Bigr ],
\end{eqnarray}
where 
\begin{equation}
\label{gc0}
\gamma =  2 v_F g_1^2/\pi^2 \hbar M R^3 .
\end{equation}
The term related to $Z_2$ is due to the stretching mode, while the term
related to $Z_3$ comes from the breathing mode.  
As pointed out above, the twisting mode  drops out in this forward-scattering
phonon exchange term, despite its importance for transport coefficients like 
the high-temperature conductivity \cite{kane}.
Notice that the retarded phonon exchange contribution
comes with a minus sign,
 indicating attractive interactions.

Further simplifications in Eq.~(\ref{finfa}) are then possible 
in the long-wavelength limit.
To lowest order in $qR$, Eq.~(\ref{coefz2}) gives $Z_3(q,\Lambda)=1$ and 
$Z_2(q,\Lambda)= \Lambda v_S^2 q^2 /(\Delta-\Lambda)\omega_B^2$, so that
\begin{eqnarray}
\frac{Z_2(q,\mu)}{\omega_n^2 + 
\epsilon_2^2(q)}  &=&    
\frac{\mu v_S^2 q^2/B\omega_B^2}{\omega_n^2 +v_S^2 q^2} ,
\label{Zstretc}\\
\frac{Z_3(q,\mu)}{\omega_n^2 + 
\epsilon_3^2(q)}  &=&    
\frac{1}{\omega_n^2 +\omega^2_B} . \label{Zbrea}
\end{eqnarray}
The forward scattering due to the stretching mode does not create a 
significant attractive interaction
since it is suppressed by a factor $(v_S/v_F)^2$ 
due to the linear dispersion of this phonon mode \cite{voit}.  A significant
attractive interaction can only be expected from the breathing mode.
Since $\omega_B$  corresponds to  room temperature for SWNT radius $R\approx 4$~\AA,
see Eq.~(\ref{matpar2}), we neglect retardation effects from the breathing mode in what
follows.  This approximation can be justified using
 the two-cutoff scheme of Ref.\cite{car}. 
Putting
 $\omega_n^2 + \omega_B^2\to \omega_B^2$ in Eq.~(\ref{Zbrea}),
interactions mediated by the breathing mode
then simply renormalize parameters in the effective LL picture.
Concerning the stretching mode, see Eq.~(\ref{Zstretc}), we make no such 
approximation, and keep its influence on the WB singularity exactly.

Straightforward algebra then leads
to exactly the action studied before in Ref.~\cite{loss},
\begin{equation} \label{finfa2}
S[\phi_{c+}] = \int [dq] \frac{ | \phi_{c+}(\omega_n,q)|^2 }{2 u_{c+}K_{c+}}
\left [ \omega_n^2 + u^2_{c+} q^2 -
\frac{b^2 q^4}{\omega_n^2 + v_S^2 q^2} \right],
\end{equation}
with the parameter identifications 
\begin{eqnarray}\label{parloss}
u^2_{c+} &=& (u^0_{c+})^2 - \gamma/\omega_B^2, \\ \nonumber
b^2 &=& \gamma \mu v_S^2 / B \omega^2_B.
\end{eqnarray}
The corresponding LL parameter $K_{c+}$ is then given by
$K_{c+} = v_F/u_{c+}$, leading directly to Eq.~(\ref{geff}) with  the radius $R_B$ 
defined as
\begin{equation}\label{rb}
R_B=\frac{2g_1^2}{\pi^2 \hbar v_F \Delta}= 2.4\pm 0.9~{\rm \AA} .
\end{equation}
The uncertainty in the estimate for $R_B$
largely results from uncertainties about the value for $g_1$, see above.
For this action (plus the standard free boson action for the neutral modes),
all correlation functions have been worked out in Ref.~\cite{loss}, and
one can simply adapt their results to the four-channel case encountered here. 

At first sight, when the radius $R$ reaches the critical value $R_c=(K^0_{c+})^2 R_B$, 
a singularity is found, which corresponds to the WB singularity
associated with the breathing phonon mode. However, due to the  additional retarded
interaction mediated by the stretching mode, i.e. the term $\sim b^2$ in Eq.~(\ref{finfa2}), 
this singularity is reached at an even larger radius. 
Following Ref.~\cite{loss}, the condition for the WB singularity is 
$b/v_S= u_{c+}$, leading to a critical radius
$R_c=(K^0_{c+})^2 R_0$ with 
\begin{equation} \label{r0}
R_0=(\Delta/B) R_B = 3.6 \pm  1.4~{\rm \AA}.
\end{equation}
The unexpectedly large value predicted for $R_0$ suggests that it may
be possible to reach this elusive singularity in practice.
This may come as a surprise, since standard 
reasoning tells us that a small ratio between sound and Fermi velocity implies 
negligible effects due to acoustic phonon exchange,
and this ratio is rather small in SWNTs.
However, the effective phonon exchange at work here
involves {\sl two} phonon branches, namely the stretching mode with $\epsilon_2(q)=v_S |q|$
and the breathing mode, $\epsilon_3(q)=\omega_B$. These two modes reinforce 
each other in driving the system towards the WB singularity.  In addition, in SWNTs one has
a very strong deformation potential that provides
very efficient electron-phonon coupling.

\begin{figure}[h]
\epsfysize=7cm 
\epsffile{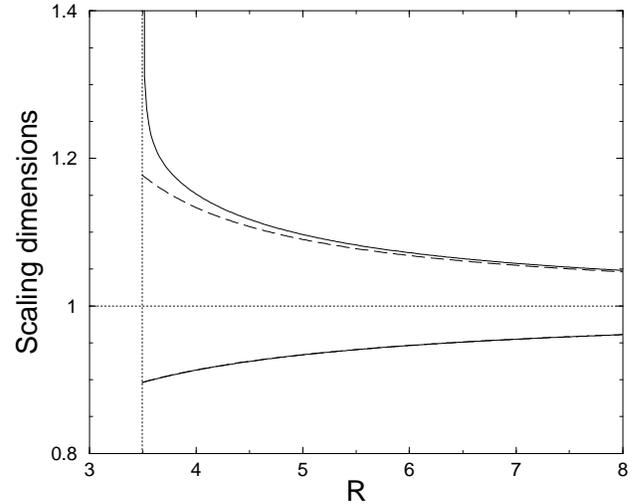} 

\caption{  \label{fig1} 
Comparison of the scaling dimensions as a function of radius $R$ (in \AA)
for SC (lower curves) and CDW (upper curves) order parameters.
Here we take the deformation potential coupling constant $g_1=25$ eV.
The dashed curves give the non-retarded LL prediction,
while the solid curves rely on Eq.~(\ref{finfa2}). The dotted vertical line 
denotes the location of the WB singularity.
In the case of the SC order parameter, the dashed and the solid curves 
cannot be distinguished.
}
\end{figure}

An important conclusion in the context of nanotube superconductivity
is that this forward-scattering phonon exchange process
is very efficient in inducing attractive electron-electron interactions.
To judge the importance of the stretching mode, we compare the scaling dimensions of 
various order parameters computed (i) under a non-retarded Luttinger liquid
description with LL parameters $u_{c+}$ and $K_{c+}$ given in Eqs.~(\ref{parloss}) and
(\ref{geff}), i.e., neglecting
the stretching mode altogether, and (ii) keeping both the breathing mode
and the retarded stretching mode, i.e. the action in Eq.~(\ref{finfa2}), 
using a straightforward generalization of exact results from Ref.~\cite{loss}.
Focussing on the scaling dimensions for charge density wave (CDW) and 
singlet superconductivity (SC) order parameters \cite{egger}, this
comparison is shown in Fig.~\ref{fig1}.  Remarkably, concerning SC there
is practically no difference at all.  For the CDW order parameter, 
a significant difference is observed only in the close vicinity  of 
the WB singularity.  
Sufficiently far away from the WB singularity, it is therefore possible to
neglect the stretching mode and describe the electronic system
as a  Luttinger liquid with (possibly) attractive interactions.
The stretching mode then only affects the location of the WB singularity but not
the magnitude of the non-retarded attractive interactions.
The LL interaction strength (including Coulomb interactions and phonon exchange)
 is determined by the parameter $K_{c+}$, 
which describes attractive interactions for $K_{c+}>1$. 
With Eq.~(\ref{geff}), we provide an estimate for $K_{c+}$ 
as a function of the tube radius $R$, the interaction strength parameter
 $K^0_{c+}$, and various material parameters.

\section{Other phonon exchange contributions and Peierls transition}\label{sec5}

Let us now consider the remaining 
terms in the phonon-mediated  action
(\ref{eleffac}).
First we combine the $J_{\alpha\pm}$ defined in Eq.~(\ref{jap}), with $\alpha=1,2=\pm$  labeling the
two K points,  as
$J_{1+} + J_{2-} = i J_{c-} + J_{CDW} $ and 
$J_{1-} + J_{2+} = - iJ_{c-} + J_{CDW}$.
Here $J_{c-} = \int\![dp] \sum_{\alpha} \alpha 
\overline \psi_\alpha  \tau_2 \psi_\alpha$ and
$J_{CDW}  =\int\![dp] \sum_{\alpha} 
\overline \psi_\alpha \tau_1 \psi_\alpha$.
The reason for these definitions becomes clear once we transform these operators
back to real space and transform from sublattice to chiral space.  
In bosonized form, 
$J_{c-}(y)= (2/\sqrt{\pi}) \ \partial_y \theta_{c-}$
and
\begin{eqnarray} \label{jbs}
J_{CDW} (y)&=& \frac{4}{\pi a} \Bigl ( \cos(2q_F y + \sqrt{\pi} \phi_{c+}) 
\cos(\sqrt{\pi} \phi_{c-}) \\ &\times& \nonumber
\sin(\sqrt{\pi} \phi_{s+}) 
\sin(\sqrt{\pi} \phi_{s-}) + [\sin \leftrightarrow \cos] \Bigr ) .
\end{eqnarray}
These expressions show that $J_{c-}$ is just the current density in the $c-$ channel,
whereas $J_{CDW}$ is the inter-sublattice 
CDW order parameter \cite{egger}.
Phonon interactions involving $J_{CDW}$ are therefore 
important in relation to the Peierls transition and describe phonon backscattering
processes.

{}From Eq.~(\ref{eleffac}), we then obtain several other contributions 
in addition to the $g_1^2$ term in Sec.~\ref{sec4},
namely $S^\prime$ and $S^{\prime\prime}$ due to the  $g_2^2$ 
and $g_1g_2$ terms, respectively. Let us start with
$S^\prime=S^\prime_{c-}+S^\prime_{CDW}$, 
where
\begin{eqnarray} \label{c-}
S^\prime_{c-} & = &- \frac{g^2_2}{L} \int [dq] \;
 | J_{c-}(\omega_n,q) |^2  \nonumber\\
&& \times 
\left[ \sin^2 (3\eta) D_+(\omega_n,q) + \cos^2 (3\eta) D_-(q)  \right],
\end{eqnarray}
and
\begin{eqnarray} \label{CDW}
S^\prime_{CDW} & = & - \frac{g^2_2}{L} \int [dq] \;
 | J_{CDW}(\omega_n,q) |^2  \nonumber \\
&&\times
\left[ \cos^2 (3\eta) D_+(\omega_n,q) + \sin^2 (3\eta)  D_-(q) \right] .
\end{eqnarray}
The inverse propagators are 
$D_-(q) =q^2/\lambda_1$ and
\begin{eqnarray}\label{dplus}
D_+(\omega_n,q) &=& 
\frac{M\omega^2_n(1+q^2R^2) + 4Bq^2 } {\lambda_2\lambda_3 R^2} 
\\ \nonumber  & =& 
\frac{1}{M R^2}\left[ \frac{Z_2(q,B)}{\omega_n^2 + \epsilon_2^2(q)} + 
\frac{Z_3(q,B)}{\omega_n^2 + \epsilon_3^2(q)} \right].
\end{eqnarray}
The functions $Z_{2,3}$ were defined in Eq.~(\ref{coefz2}).
In principle, there is another contribution to $S^\prime$ from mixed terms 
involving the product of $J_{c-}$ and $J_{CDW}$. However, such terms vanish for a 
doped SWNT by virtue of momentum conservation. In real space, this is reflected
by the oscillatory $\cos(2q_F y)$ factor appearing in $J_{CDW}$.  Since this factor
is absent in $J_{c-}$, we  ignore the mixed terms in what follows.
The second contribution to the action reads
\begin{eqnarray} \label{g1g2}
S^{\prime\prime} &=& \frac{g_1g_2}{L}
\sin (3\eta) \int [dq]\\ \nonumber &\times&  \rho_{c+}(-\omega_n,-q) D_m(\omega_n,q) 
J_{c-}(\omega_n,q) ,
\end{eqnarray}
with
\begin{eqnarray} \label{dm}
D_m(\omega_n,q) &=& 
\frac{M\omega^2_n (1-q^2R^2) }{\lambda_2 \lambda_3 R^2}
\\ \nonumber & = & \frac{1}{M R^2}
\left[ \frac{\tilde Z_2(q)}{\omega_n^2 + \epsilon_2^2(q)} + 
\frac{\tilde Z_3(q)}{\omega_n^2 + \epsilon_3^2(q)} \right] ~,
\end{eqnarray}
where
$\tilde Z_{2,3}(q) =\mp (1-q^2R^2) M R^2 \epsilon^2_{2,3}(q)/\Delta E(q)$,
with $E(q)$ given in Eq.~(\ref{defb}).
At long wavelengths, this yields
$\tilde Z_2(q)= -v^2_S q^2/\omega_B^2$ and
 $\tilde Z_3(q) = 1$. Then Eq.~(\ref{dm}) simplifies to
\begin{equation}\label{dm2}
D_m(\omega_n,q\ll R^{-1}) = 
\frac{\omega_n^2/\Delta}{\omega_n^2 + v_S^2q^2}.
\end{equation}
Again, there is in principle another contribution that arises from mixing 
$\rho_{c+}$ with $J_{CDW}$.  However, this contribution
also vanishes by virtue of momentum conservation and is disregarded.

Next, we analyze these additional terms using bosonization, starting with 
the forward-scattering term $S^\prime_{c-}$,
which can be treated along the same lines as the 
forward-scattering term in the $c+$ channel, see Sec.~\ref{sec4}.
Neglecting retardation in the breathing mode as above, 
the long-wavelength 
action for the bosonic field $\theta_{c-}$ including $S^\prime_{c-}$ is
\begin{eqnarray}
\label{totalactc-}
S[\theta_{c-}] &=& \frac{1}{2v_F} \int \! [dq] 
| \theta_{c-}(\omega_n,q) |^2 
\Bigl \{ \omega_n^2 + v^2_{F} q^2  
- \gamma' q^2  
 \\ \nonumber &\times&
\Bigl [ \sin^2 (3\eta) \left( 
		\frac{ 1 }{\omega^2_B } + 
		\frac{ 4B^2q^2R^2/\Delta^2 }{ \omega_n^2 + v_S^2 q^2 } 
			\right) \\ \nonumber
 &+& \cos^2 (3\eta) \frac{q^2R^2}{\omega^2_n + v^2_T q^2} \Bigr] \Bigr \},
\end{eqnarray}
with the parameter
$\gamma'=  2 v_F g_2^2 /\pi^2 \hbar M R^3 $.
Amusingly, the contributions of different phonon modes are
weighted by chirality-dependent factors. In particular, for zig-zag tubes
($\eta=0$) only the twisting mode contributes, whereas for armchair tubes
($\eta=\pi/6$) only breathing and stretching modes do. 
Again we obtain a WB singularity but now in the $c-$ channel 
at a corresponding critical radius
$R^\prime_0 =2g^2_2/\pi^2 \hbar v_F \mu$.  Inserting the parameter values
given above, due to the small ratio $g_2/g_1$, the critical radius 
is exceedingly small,
$R^\prime_0 \approx 0.02$~\AA.  Using the same reasoning as in Sec.~\ref{sec4},
 the effect of $S_{c-}^\prime$ again leads to a LL
action in the $c-$ channel, with a renormalization 
of $u_{c-}$ and $K_{c-}$.
However, such renormalizations  amount to tiny changes of the order $10^{-3}$
 even for very thin SWNTs. Therefore $S_{c-}^\prime$ does not imply observable
consequences and is omitted henceforth.

\begin{figure} 
\epsfysize=7cm 
\epsffile{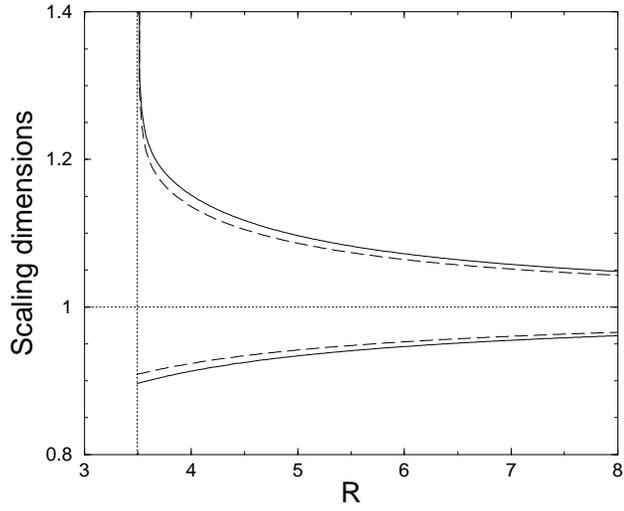}
 
\caption{ \label{fig2} 
Comparison of the scaling dimensions as a function of radius $R$ (in\AA) 
for the SC (lower curve) and the CDW (upper curve) order parameter, 
computed with $\epsilon=0$
(dashed curves) and $\epsilon =0.06$ (solid curves). 
The calculation employs  Eq.~(\ref{finfa2}). 
Taking $g_1=25$ eV, the value $\epsilon=0.06$ corresponds 
to $g_2 = 1.5$ eV.
We consider the case of an armchair tube ($\eta=\pi/6$), 
where the effect of $S^{\prime\prime}$ is most
pronounced.
}
\end{figure}

Next we focus on $S^{\prime\prime}$, which mixes the $c+$ and $c-$ sectors and
constitutes another forward scattering mechanism.
First, we observe that this term vanishes for zig-zag tubes ($\eta=0$).
Neglecting again retardation in the breathing mode and keeping only terms to lowest
order in $q R$, we find from Eq.~(\ref{dm2}) 
\begin{equation}\label{s2}
S^{\prime\prime} = \epsilon \kappa
\int [dq] \phi_{c+}(-\omega_n,-q) \theta_{c-}(\omega_n,q) 
 \frac{\omega_n^2 q^2} {\omega_n^2+v^2_S q^2} ,
\end{equation}
with $\epsilon= g_2/g_1$ and $\kappa =\sin(3\eta) v_{F} R_B/ \hbar R$.
Introducing the notation $\Phi^T = (\phi_{c+},\theta_{c+},\phi_{c-},\theta_{c-})$,
the total action [including Eq.~(\ref{finfa2}) and
$S^{\prime\prime}$] can be written as
\begin{equation}\label{tott}
S=\frac12 \int \! [dq] \Phi^T  [ \underline{S}_0  + 
\epsilon \underline{S}_1 ] \Phi,
\end{equation}
where $\underline {S}_0$ is a block-diagonal matrix in $(c+,c-)$ space, and 
$\underline{S}_1$ contains the mixing term due to $S^{\prime\prime}$.
Since $\epsilon \ll 1 $, 
we proceed by expanding to first order in $\epsilon$.
This allows the explicit calculation of scaling dimensions of all 
order parameters of interest.
Details about this calculation are contained in an
appendix, and the main results are shown in Fig.~\ref{fig2}.  
Clearly, the effect of $S^{\prime\prime}$ is also very tiny, even for
very thin tubes.  Neglecting this term, only the term $S^\prime_{CDW}$ describing
backscattering phonon exchange remains. 

Let us then study the $S^\prime_{CDW}$ term. 
As this describes an effective interaction involving $J_{CDW}$, this
part monitors the possibility of a Peierls transition.
In order to keep the discussion transparent, we focus on
 the static limit. Then Eq.~(\ref{CDW}) reduces to  
\begin{equation}\label{cdw2}
S^\prime_{CDW} =  - \frac{g^2_2}{\mu L} \int [dq] \
 | J_{CDW}(\omega_n,q) |^2 .
\end{equation}
Transforming to real space, using the bosonized expression of $J_{CDW}$ in Eq.~(\ref{jbs}),
we find a term involving $4q_F$ oscillatory contributions that violate momentum
conservation and average to zero, and in addition a tiny renormalization of the
LL parameters $u_\lambda$ and $K_\lambda$. 
This renormalization represents a direct generalization of the one/two-channel theory
of Voit and Schulz \cite{voit} to the four-channel case of interest here.
We do not give any details here, since the relevant dimensionless coupling 
constant is $R^\prime_0/\pi^3 R$,  and the resulting renormalizations 
are again of order $10^{-3}$,
i.e.  unobservable. 
Nevertheless, we confirm the finding of Ref.\cite{peierls} that the
relevant Peierls distortion mode for armchair SWNTs is given by a twist distortion,
see Eq.~(\ref{CDW}). For arbitrary chiral angle $\eta$, however, the relevant
distortion involves a linear combination of all three phonon modes.
Finally, we mention that on half-filling, the oscillatory contribution could
survive.  However, its scaling dimension suggests that this term is strongly
irrelevant unless electron-electron backscattering gaps out the neutral channels.
For very thin tubes, these gaps may be sufficiently pronounced \cite{egger} to allow for 
 Eq.~(\ref{cdw2}) to be a relevant perturbation. Provided $K_{c+}<2$, one could possibly
expect a small (twist) Peierls gap in this nearly commensurate situation,
see Ref.\cite{peierls}.  
However, this would require a detailed analysis of additional phonon exchange processes
neglected above due to incommensurability.
Away from half-filling,  we do not see a possibility for this gap to survive.

\section{Discussion and conclusions} \label{sec7}

In this paper, we have studied acoustic
phonon exchange mechanisms operating in an individual metallic
SWNT.  The description of phonons within an elastic continuum
model allows for explicit analytical progress, and we have shown that 
integrating out the phonon system can produce very substantial attractive
electron-electron interactions.  For sufficiently large tube radius,
the system can be described by a standard Luttinger liquid theory, where
the interaction parameter $K_{c+}$ is renormalized away from the value 
$K_{c+}^0\leq 1$ describing the theory without phonons but fully taking
into account Coulomb interactions.   We have
shown that $K_{c+}>1$ corresponding to attractive interactions is possible,
in particular once the Coulomb interactions are externally screened.
 These attractive interactions are basically mediated via a breathing phonon
mode.  For thin SWNTs, one also has to take into account a retarded
branch describing a stretching mode.  The combined effect of the stretching
and the breathing mode then produces a Wentzel-Bardeen singularity at a critical
radius $R_0\approx 3.6$~\AA.  Nanotubes may then offer the possibility
to reach this singularity in experiments.  Remarkably,  all other
mechanisms, e.g. backscattering phonon exchange, turn out to be remarkable
inefficient in mediating effective electron-electron interactions.
In particular, based on our analysis, it should be extraordinarily difficult
to observe traces of the Peierls  transition.  As our treatment is essentially
exact, we believe that this also invalidates some earlier theories that were
mainly based on mean-field approximations. 

As mentioned in the introduction, the phonon-mediated attractive interaction
can only drive the system into a true superconducting phase if  
inter-tube couplings are taken into account. This will be studied in detail
elsewhere \cite{future}.
In order to gain preliminary insights  about inter-tube phonon exchange effects in
a rope of SWNTs or a multi-wall nanotube,
let us briefly contemplate a simplified model  of two parallel SWNTs and 
study phonon exchange in the presence of elastic couplings between both tubes.
We shall focus on the breathing mode alone,  which will produce
the most important effects. In the long-wavelength limit, it 
can be approximated as a dispersionless mode,
leading to the free-phonon action
\[
S = \frac{M}{2L} \int [dq] \sum_{i,j=1,2} u_i(-\omega_n,-q) A_{ij}(\omega_n) u_j(\omega_n,q) ,
\]
where $u_i$ is the radial component ($u_z$) of the displacement field on tube $i$, 
\[
A_{ij}(\omega_n) =
\left( \begin{array}{cc}
\omega^2_n + \omega^2_B(1 + \alpha_l) & -  \alpha_t \omega^2_B \\
-  \alpha_t \omega^2_B & \omega^2_n + \omega^2_B(1 + \alpha_l)
\end{array}
\right) ,
\]
and $\alpha_t,\alpha_l$ are dimensionless parameters.  While one can compute
these parameters from a microscopic force-constant model, here we take
them simply as phenomenological parameters.  On general grounds, they must 
obey $0\leq \alpha_{t,l} \leq 1$, where $\alpha_{t,l}=0$ in the absence of 
inter-tube couplings.  With total electronic densities $\rho_i$ on tube $i=1,2$, 
the dominant electron-phonon interaction is due to the deformation potential (\ref{vi}),
producing a contribution
\[
(g_1/2\pi R^2) \int [dq] \sum_{i=1,2} \rho_i(-\omega_n,-q) u_i(\omega_n,q) 
\]
to the action.
We can diagonalize this action by introducing symmetric and antisymmetric 
combinations, $u_{s/a}=(u_1 \pm u_2)/\sqrt{2}$ and 
$\rho_{s/a}=(\rho_1 \pm \rho_2)/\sqrt{2}$. Integrating out the phonons,
the forward-scattering contribution in the $c+$ channel 
leads to a Luttinger liquid,  but with 
renormalized parameters,
\begin{equation} \label{sc+}
S[c+]= \frac12 \sum_{j=s,a} \int [dq] \frac{|\phi_{j,c+}|^2} {u_{j,c+} K_{j,c+}}
\left[ \omega_n^2 + (u_{j,c+})^2 q^2 \right] ,
\end{equation}
where $u_{s/a,c+}  =  v_F/K_{s/a,c+}$. The interaction parameters are
\begin{equation}\label{kcps}
K_{s/a,c+} =  \frac{ K_{c+}^0 } { \sqrt{1-(K_{c+}^0)^2 R_B/ f_{s/a} R } } ,
\end{equation}
where $f_{s/a} = 1 + \alpha_l\mp\alpha_t$.

There are now two types of superconducting order parameters.  
Focusing on the most important singlet superconducting fluctuations, there is
one operator describing
a Cooper pair on the same tube, 
${\cal O}^{\mbox{\tiny intra}}_{SSC0}$,
 and another one where the electrons forming the
Cooper pair reside on different tubes,
${\cal O}^{\mbox{\tiny inter}}_{SSC0}$, with explicit form 
\begin{eqnarray*}
{\cal O}^{\mbox{\tiny intra}}_{SSC0} & 
\sim & \sum_{r\alpha\sigma} \sigma \psi_{1,r\alpha\sigma} 
\psi_{1,-r-\alpha -\sigma} , \\
{\cal O}^{\mbox{\tiny inter}}_{SSC0} & \sim & 
\sum_{r\alpha \sigma } \sigma \psi_{1,r\alpha\sigma} 
\psi_{2,-r-\alpha -\sigma} ,
\end{eqnarray*}
where $\psi_{1/2,r\alpha\sigma}$ is the electron spinor on tube $1/2.$
{}From Eq.~(\ref{sc+}), the scaling dimensions of these 
operators follow as
\begin{eqnarray}
\Delta^{\mbox{\tiny intra}}_{SSC0} & = & 
\frac{1}{8} \left( 6 + \frac{1}{K_{s,c+}} +\frac{1}{K_{a,c+}} \right), \\
\Delta^{\mbox{\tiny inter}}_{SSC0} & = & \frac{1}{8} \left( 6 + \frac{1}{K_{s,c+}} + 
K_{a,c+} \right). 
\end{eqnarray}
Two conclusions can be drawn from the inspection of these scaling dimensions. 
First, for inter-tube couplings  with $\alpha_t$ sufficiently larger than $\alpha_l$,
attractive interactions within a given tube are increased. Second,
whenever $K_{a,c+} >1$, which holds true in the absence of Coulomb
interactions, the intra-tube superconducting order parameter is always 
more relevant than the inter-tube one.  In other words, we expect that
{\sl Cooper pairs predominantly form on a given tube} but not across 
different tubes.  Remarkably, this conclusion does not depend on
the values of $\alpha_{l,t}$.  Clearly, this finding will be important in the
theoretical analysis of experiments on SWNT ropes and related systems. 

Finally, we briefly address the role of electron-electron backscattering
that has been neglected in our study.  {}From Ref.~\cite{egger}, backscattering
is expected to open tiny gaps that depend exponentially on the radius $R$.
While for $R>5$~\AA, the gap is extremely small and 
of no relevance in practice, for ultrathin tubes ($R\approx 2$~\AA) 
the gaps may be more important.  However, in this limit also the bandstructure may
change in a profound way due to  hybridization of $\pi$ and $\sigma$ orbitals, and the
analysis of Ref.\cite{egger} cannot simply be taken over.  In addition, the
presence of the Wentzel-Bardeen singularity poses an intrinsic 
lower limit to the validity and self-consistency of
our effective low-energy approach. With the predicted value
$R_0\approx 3.6$~\AA, we therefore do not expect 
that backscattering leads to dramatic changes to the picture put forth in this paper.

\acknowledgements
This work has been supported by the EU and by the DFG.

\appendix
\section{Perturbative treatment of $S^\prime$}

In this appendix, we briefly describe some details concerning the perturbative treatment
of $S^\prime$, see Eq.~(\ref{s2}), in Sec.~\ref{sec5}.  With the notation
$\Phi^T = (\phi_{c+},\theta_{c+},\phi_{c-},\theta_{c-})$, the full action in the
total/relative charge channels is Eq.~(\ref{tott}) with the block-diagonal matrix 
$\underline{S}_0={\rm diag}(\underline{A}_{c+},\underline{A}_{c-})$,
where
\begin{eqnarray*} 
\underline{A}_{c+} &=&
\left( \begin{array}{cc}
\frac{u^2_{c+}}{v_F}q^2 - \frac{b^2q^4/v_F}
{\omega_n^2+v_S^2 q^2}& - i\omega_n q  \\
- i\omega_n q  & v_F q^2
\end{array}
\right), \\
\underline{A}_{c-} &=&
\left( \begin{array}{cc}
v_F q^2 & - i \omega_n q \\
- i \omega_n q & v_F q^2
\end{array}
\right).
\end{eqnarray*}
Furthermore, the mixing term leads to a matrix $\underline{S}_1$, whose only
non-zero entry is
\[
(\underline{S}_1)_{14} = (\underline{S}_1)_{41} = \kappa \omega_n^2 q^2/(\omega_n^2
+ v_S^2 q^2) ,
\]
where $\kappa$ is defined immediately after Eq.~(\ref{s2}).
We then expand to first order in $\epsilon$,
\begin{eqnarray*}
(\underline S_0 + \epsilon \underline S_1)^{-1} &=&
 \underline S^{-1}_0 - \epsilon \underline S^{-1}_0\underline S_1 
\underline S^{-1}_0 + {\cal O}(\epsilon^2)
\\ &=& S^{-1}_0 - \epsilon \kappa \left( 
\begin{array}{cc}
0 & \underline{H}\\ \underline{H}^T & 0 \end{array}
\right).
\end{eqnarray*}
Some algebra yields for the matrix $\underline{H}$ the following result: 
\[
\underline H = 
\left( 
\begin{array}{cc}
(i v_F  \omega_n/q) {\cal P} & 
  v^2_F {\cal P}\\
{\cal P}^\prime & (iv_F\omega_n/q) {\cal P}
\end{array}
\right),
\]
where
\[
{\cal P} = \sum_{\beta=\pm,i} \frac{F_\beta Q^\beta_i}{\omega^2_n + 
v_{i\beta}^2 q^2}, \quad 
{\cal P}^\prime = \sum_{\beta=\pm,i} \frac{ v^2_{i\beta} 
F_\beta Q^\beta_i}{\omega^2_n + v_{i\beta}^2q^2} .
\]
Here $i=0,1,2$ and we denoted $v_{1\pm}=v_S, v_{2\pm}=v_F$. The 
velocities $v_{0\pm}\equiv v_{\pm}$ and the coefficients $F_\beta, C_\beta$ 
are given in Ref.~\cite{loss}:
\begin{eqnarray*}
2 v^2_{\pm} &=& u_{c+}^2 + v_S^2 \mp \sqrt{(u^2_{c+}-v_S^2)^2 + 4 b^2} , \\
F_\beta &=& \frac{v_\beta^2 -v^2_S}{v^2_\beta - v^2_{-\beta}} , \\
C_\beta &=& \frac{u_{c+}}{v_\beta}\frac{v_\beta^2-v_S^2 + b^2/u_{c+}^2}
{v^2_\beta - v^2_{-\beta}} .
\end{eqnarray*}
Finally, the coefficients $Q^\beta_i$ are defined by
\begin{eqnarray*}
Q^\pm_0 &=& \frac{- v^2_{\pm}}{(v_{\pm}^2 - v_S^2)(v_{\pm}^2-v_F^2)} , \\
Q^\pm_1 &=& \frac{- v^2_S}{(v_S^2-v_F^2)(v_S^2-v^2_{\pm})} ,\\
Q^\pm_{2} &=& \frac{- v^2_F}{(v_F^2-v_S^2)(v_F^2-v^2_{\pm})} .
\end{eqnarray*}
With this, it is straightforward to compute correlation functions for the 
bosonic field.
The $T=0$ equal-time correlation functions $\langle \Phi_i(x) \Phi_j(0)\rangle$
are given by 
\[
\langle \Phi_i(x) \Phi_j(0)\rangle = -\frac{G_{ij}}{4\pi} \ln ([x^2+a^2]/L_0^2),
\]
where $a$ and $L_0$ are UV and IR cutoff lengths. The only non-zero entries
of $G_{ij}$ are $G_{33}=G_{44}=1$ and
\begin{eqnarray*}
G_{11} &=& \sum_{\beta} v_F F_\beta/v_\beta ,\\
G_{14} &=& -\epsilon \kappa v^2 _F \sum_{\beta,i}F_{\beta} Q^\beta_i/v_{i\beta} ,\\
G_{22} &=& \sum_{\beta} u_{c+} C_\beta/v_F ,\\
G_{23} &=& -\epsilon \kappa \sum_{\beta,i} v_{i \beta} F_{\beta} Q^\beta_i .
\end{eqnarray*}
With this,  the scaling dimension of  
any operator $\exp [i\alpha^T \Phi(x)]$ is found as
$\alpha^T  \underline{G} \alpha /4\pi$.

\end{document}